\documentclass[aps,showpacs,prd,twocolumn]{revtex4}
\usepackage{graphicx}
\usepackage{amssymb}
\usepackage{amsmath}
\newcommand{\be}{\begin{equation}}
\newcommand{\ee}{\end{equation}}
\newcommand{\ben}{\begin{eqnarray}}
\newcommand{\een}{\end{eqnarray}}
\newcommand{\bes}{\begin{subequations}}
\newcommand{\ees}{\end{subequations}}
\newcommand{\bb}{\bibitem}

\begin{document}
\title{Getting inflationary models using the deformation method}
\author{J.J. Rodrigues}
\affiliation{Universidade Estadual da Para\'{\i}ba 
58233-000 Araruna, Para\'{\i}ba, Brasil}
\author{M.A.M. Souza}
\affiliation{Instituto Federal de Educa\c c\~ao, Ci\^encia e Tecnologia do Piau\'i 
64215-000 Parna\'iba, Piau\'i, Brasil}

\date{\today}\date{\today}

\begin{abstract}
We show as the dynamics for the inflaton, under slow-roll regime, can be treated in a other dynamics, following the deformation procedure. In a direct way we present a relationship between two slow-roll inflationary potentials, and we apply this framework to show how to construct an eternal inflation from chaotic inflation, or even, a natural inflation from hilltop inflation, easily.

\end{abstract}

\maketitle

\section{Introduction}

The early universe is full of challenging problems that remain to fill the tables of many high energy theoretical physics. Several of these problems can be addressed through the inflation theory \cite{ITheory,starobinsky,linde}, where a scalar field, called inflaton, can be invoked to drive the evolution of the early universe. Particularly challenging is the choice of a dynamics to the inflaton, provided that a large number of parameters should be adjusted to ensure the success of the model and then bring the universe to graceful exit \cite{GExit,steinhardt,veneziano,brown}. Following this line, theoretical developments to simplify and put the inflation models within the observational constraints are welcome.  

In this paper we introduce a framework to implement a correlation between two dynamics for the inflaton under the slow-roll regime \cite{SRoll,leach,martin} (see also references therein), where the inflation parameters present a direct relationship with a potential that drives the solution. Especially, we show that the choice of dynamics to inflaton can be treated in another correspondence model, and this is based in the deformation procedure \cite{DPFT,bazeia,bazeiaD,gonzalez,afonso}.

The work is organized as follows: Sec. \ref{sec2} presents the generalities about the deformation procedure in field theory. The cosmological background is described in Sec. \ref{sec3}, where we established that the inflaton following a slow-roll dynamics. In the Sec. \ref{sec4} we apply the deformation procedure to show as a new inflation solution can be studied, from a known inflation solution; being the Sec. \ref{sec5} dedicated to show just a few examples, that illustrate this framework. The comments are presented in the Sec. \ref{sec6}.

Throughout this work we use units in which $c={4\pi G}=H_0=1$, where $c$ is the speed of light in vacuum, $G$ is the gravitational constant, $H$ is the Hubble parameter and the subscript `0' refers to the initial time.

\section{Deformation procedure in field theory} \label{sec2}

A result of advances in research in high energy physics, an innumerable class of models described by a scalar field has been proposed. However the difficulty lies in the fact that many of these models do not provide an analytical description of the system studied, which hinders the complete understanding of these systems. It is therefore necessary to find a method that can generate potentials with the analytical solutions and of physical interest. A rather effective method is the so-called deformed procedure, proposed by Bazeia et al, consisting of generating new solutions from a potential, or a solution, of a known model, with the aid of a properly deformation function.

The advantage of the method is the fact that the description of the characteristics of the new model can be done analytically, without having to resort to computational methods, or numerical analysis. The relationship between the potential of the original model $V(\chi)$, and the potential of the obtained model, $V(\phi)$, is given for
\be
V(\phi)=\frac{V(\chi\to f(\phi))}{f^{2}_{\phi}(\phi)}
\ee
where $f(\phi)$ is the deformation function. In this case, if $\chi(x)$ is a static solution of the starting model, then we get that 
\be
\phi(x)=f^{-1}(\chi(x))
\ee
where $\phi(x)$ is a solution of the new deformed model. We can still see that for the case of topological solutions a deformed defect $\phi(x)$ connects the corresponding minimums of the solutions $\chi(x)$ of the original model, given for $\tilde{\upsilon}_{i}=f^{-1}(\upsilon_{i}), i=1,2,3,...,n$.

\section{Cosmological background} \label{sec3}

We shall consider a model in which the dynamics of the early universe is described by the Einstein-Hilbert action, where a scalar field $\chi$ is minimally coupled to gravity, i.e.
\be
S=\int\,d^4x\;{\sqrt{-g}\;\left(-\frac14\,R+{\mathcal L(\chi,X)}\right)}
\ee
$R$ is a scalar curvature, being ${\mathcal L(\chi,X)}$ the lagrangians scalar field with $X=\chi_{,\mu}  \chi^{,\mu}/2$ and a comma represents a partial derivative. In the following it will be assumed that $\chi$ plays the role of the inflaton field.

Consider a flat Friedmann-Robertson-Walker background with line element
\be
ds^2=dt^2 - a^2(t)\left(dx^2+dy^2 +dz^2\right) \,,
\ee
where $t$ is the physical time and $x$, $y$ and $z$ are comoving spacial coordinates.

We assume that the energy-momentum tensor for the inflaton field can be written in a perfect fluid form
\be\label{eq:fluid}
T^{\mu\nu}= (\rho+ p) u^\mu u^\nu - p g^{\mu\nu} \,,
\ee
by means of the following identifications
\be
u_\mu = \frac{\chi_{, \mu}}{\sqrt{2X}} \,,  \quad \rho = 2 X {\mathcal L}_{,X} - {\mathcal L} \, ,\quad p =  {\mathcal L}(X,\chi)\,.
\ee
In Eq.~(\ref {eq:fluid}), $u^\mu$ is the 4-velocity field describing the motion of the fluid (for timelike $\chi_{, \mu}$), while $\rho$ and $p$ are its proper energy density and pressure, respectively.

Solving the action to the metric above the Einstein equations reduce to 
\be
H^{2}=\frac{2}{3}\,\rho
\ee
and
\be
\frac{\ddot{a}}{a}=-\frac{1}{3}(\rho+3p)
\ee
where $H={\dot a}/a$ and dot represents a derivative with respect to physical time.

If we consider the standard dynamics, described by lagrangian density
\be
{\mathcal L}=\frac12 \chi_{,\mu} \chi^{,\mu}-V(\chi)
\ee
the continuity equation can be written as
\be\label{me1}
\ddot{\chi}+3H\dot{\chi}+V_{\chi}=0
\ee
the index $_{\chi}$ presents a derivative related to the field. The pressure and energy density are given by
\be
\rho=\frac12\dot{\chi}^2+V,\;\;\;\ p=\frac12\dot{\chi}^2-V
\ee
In this way, we can rewrite the solutions of Einstein equations. We obtain
\be\label{fe1}
H^{2}=\frac{1}{3}\dot{\chi^{2}}+\frac{2}{3}V
\ee
and
\be\label{fe2}
\dot{H}=-\dot{\chi^{2}}
\ee

Inflationary solutions for which the energy density of the universe is dominated by the potential term $V(\chi)$ enable us choose a known approach as slow-roll approximation, where the inflaton does not vary too rapidly and we can neglect the kinetic term in the Friedmann equation and the acceleration term in the equation of motion of the scalar field, which leads us naturally to first-order equations, such as
\be
H^{2}\approx\frac{2}{3}V(\chi)
\ee
\be
3H\dot{\chi}+V_{\chi}\approx0
\ee
These equations show that the choice of the potential allows us to apply limits to the inflationary parameters. The number N of e-fold, written as $N=\ln(a_\text{end}/a)$, where $a_\text{end}$ is the scalar factor in the end of inflation, can be obtained, it is given that 
\be\label{av}
a=a_0\,\exp\left(\int_{t_i}^{t_e}\left(\frac23\,V(\chi)\right)^{1/2}dt\right)
\ee 
i.e. $N=\int^{t_\text{end}}_{t}H\,dt$, being the inflaton evolution determined for \eqref{av}. In a similar way, to establish the flatness condition, we need that the slow-roll parameters, defined as \cite{SRI,lyth}
\be
\epsilon=\frac14\left(\frac{V_{\chi}}{V}\right)^2,\;\;\;\;\;\eta=\frac12\frac{V_{\chi\chi}}{V}
\ee
are such that $\left|\epsilon\right|<<1$, as well as, $\left|\eta\right|<<1$, and all parameters are sensible the choice of the potentials. 

The deformation procedure creates a class of the analytical potentials and this can be an easy way to analyze these parameters, or even, all parameters that are functions of the potential. We are going to see now as this procedure can be applied to slow-roll inflation scenario.  

\section{Deforming slow-roll inflationary models} \label{sec4}

Initially we consider that the inflaton has its dynamics described by the lagrangian density
\be
{\cal L}=\frac12\chi_{,\mu} \chi^{,\mu}-V(\chi)
\ee
where $V(\chi)$ presents the potential field. The continuity equation in this dynamics takes the form 
\be\label{ce1}
\rho_{\chi}+3H\dot{\chi}=0
\ee
Since we know that
\be
H^2=\frac23\rho
\ee
and squared \eqref{ce1} we obtain the useful relation
\be\label{rhoc}
6\rho{\dot{\chi}}^2=\rho_{\chi}^{2}
\ee
Now we consider another dynamics for the inflaton evolution described by the lagrangian density
\be
{\cal L}=\frac12\phi_{,\mu} \phi^{,\mu}-\tilde{V}(\phi)
\ee
Similarly to the previous model results
\be\label{rhop}
6\tilde{\rho}{\dot{\phi}}^2=\tilde{\rho}_{\phi}^2
\ee

The key point of this description is to redefine the dynamics field  via the relation
\be
\chi=f(\phi)
\ee
where $f(\phi)$ is a so called deformation function. As a direct consequence of this definition we can write 
\be
\dot{\phi}=\frac{\dot{\chi}}{f_{\phi}}
\ee
in which $f_{\phi}=df/d\phi$. Using \eqref{rhoc}, and \eqref{rhop} we come to 
\be\label{ded}
\frac{\tilde{\rho}_{\phi}^2}{\tilde{\rho}}=\frac1{f_{\phi}^2}\left(\frac{\rho_{\chi}^2}{\rho}\right)_{\chi=f(\phi)}
\ee
this presents a generic correspondence between two energy densities describing two dynamics scalar field. This is a natural way to deform the energy density for two fluids in the same background provided that these fluids have a conserved energy-momentum tensor, or better, provided that they satisfy a continuity equation. 

The slow-roll condition applied to the equation of motion of the scalar field allows us to rewrite it as 
\be
3H\dot{\chi}=-V_{\chi}
\ee
now
\be
6V{\dot{\chi}}^2=V_{\chi}^{2}
\ee
as well as
\be
6\tilde{V}{\dot{\phi}}^2=\tilde{V}_{\phi}^2
\ee
and we have
\be\label{defor}
\frac{\tilde{V}_{\phi}^2}{\tilde{V}}=\frac1{f_{\phi}^2}\left(\frac{V_{\chi}^2}{V}\right)_{\chi=f(\phi)}
\ee
this presents a generic correspondence between two potentials describing the inflaton dynamics, under slow-roll approximation. 

The solutions in the new model are obtained with $\phi=f^{-1}(\chi)$, this is the inverse deformation function calculated with the solutions of original model. An important implication of this framework is based under the possibility of obtaining an analytical description for new inflation solutions and allows to analyze the parameters for these solutions, provided we have known the parameters for the original inflationary model, minimizing or even annulling the numerical techniques, making the search for more complicated vacuum configurations more accessible.

The limit in that slow-roll condition ceases to be valid $\dot{\chi}^2/2=V(\chi)$ depends on the potential chosen and we have
\be
t_{end}-t_{ini}=\int^{\chi_{end}}_{\chi_{ini}}\frac{d\chi}{\sqrt{2V}}
\ee
which leads to $t_{end}-t_{ini}\,\,\neq\,\,\tilde{t}_{end}-\tilde{t}_{ini}$, since that \eqref{defor} is a valid relation between the potential of the original model and the potential of the deformed model. In this sense the deformation procedure does not show the correlation between two slow-roll sectors, but between two potentials that have a slow-roll regime by construction, which can be seen by analyzing the deformation of the slow-roll parameters.

\section{Aplications} \label{sec5}

To illustrate this framework initially we deal with a model based in the chaotic inflation \cite{CI,lindeA,wands}. In this model the dynamics field is driven by the quadratic potential $V(\chi)=V_0\chi^2$. The deformation procedure can lead us directly to an eternal inflation model \cite{EI,vilenkin,aguth}, described by potential $\tilde{V}(\phi)=\tilde{V}_0\phi^p$, where we choose $p>2$, assuming the deformation function 
\be
f(\phi)=\chi=-4\sqrt{\frac{{V}_0}{\tilde{V}_{0}}}\frac{\phi^{-\frac12(p-4)}}{p(p-4)}\,,
\ee
applied to potential chaotic inflation and using \eqref{defor}.

Once the potentials are known we can obtain the slow-roll parameters. To the original potential we have $\epsilon=\eta={\chi^{-2}}$. It applying the deformation procedure, these parameters are obtained in the other frame and now we write
\be
{\tilde{\epsilon}}=\frac{p^2}{4\phi^2},\;\;\;\;\;\;\;\;{\tilde{\eta}}=\frac{p(p-1)}{2\phi^2}
\ee
The end of inflation occurs now with an additional choice of parameter $p$ for the deformed model, while only conditions for fields are necessary for the original model.  

The e-fold number can be estimated in this framework as
\be
N=\frac12(\chi^2_{\text{end}}-\chi^2_{\text{ini}}) 
\ee
and to the deformed model
\be
\tilde{N}=\frac1{p}(\phi^2_{\text{end}}-\phi^2_{\text{ini}}),
\ee
We note that the case $p=4$ must be analyzed separately. To this choice, we write $\tilde{V}(\phi)=\lambda\phi^4$ and following the previous results, the deformation function is
\be
f(\phi)=\chi=\sqrt{\frac{V_0}{\lambda}}\frac{\ln{\phi}}2
\ee
Now the slow-roll parameters are
\be
\tilde{\epsilon}=\frac4{\phi^2},\;\;\;\;\;\;\;\;\tilde{\eta}=\frac6{\phi^2}
\ee
To this case the e-fold number is given by same $\tilde{N}$ for $p=4$.

We now consider as the initial model the hilltop inflation \cite{HI,lin,pal}, being
\be
V(\chi)=\left(V_0-\frac{\lambda}{p}\,\chi^{p/2}\right)^2
\ee
This potential can be deformed directly in the natural inflation potential \cite{NI,adams,randall}, given by $\tilde{V}(\phi)=\tilde{V}_0\cos^2(r\phi)$, when we choose the following deformation function
\be
f(\phi)=\chi=\left(\frac{4\sqrt{\tilde{V}_0}r^2}{(p-4)\,\lambda\,{\rm arctanh}(\cos(r\phi))}\right)^{2/({p-4})}
\ee
being the integration constant such that $f(\phi=\pi/(2r))=0$.

Here, the slow-roll parameters for the original potential are
\ben
\epsilon&=&\frac14\frac{\chi^{p-2}}{\displaystyle\left(\frac{\chi^{p/2}}{p}-\frac{V_0}{\lambda}\right)^2}
\\
\eta&=&\frac{\displaystyle\frac{p-1}{2p}\chi^{p-2}-\frac14\frac{V_0}{\lambda}(p-2)\chi^{(p-4)/2}}{\displaystyle\left(\frac{\chi^{p/2}}{p}-\frac{V_0}{\lambda}\right)^2}
\een
In the other way, for the deformation potential we have
\be
{\tilde{\epsilon}}=r^2\tan^2(r\phi),\;\;\;\;\;\;\;\;{\tilde{\eta}}=r^2(\tan^2(r\phi)-1)
\ee

\begin{figure}[htb!]
\includegraphics[scale=.35]{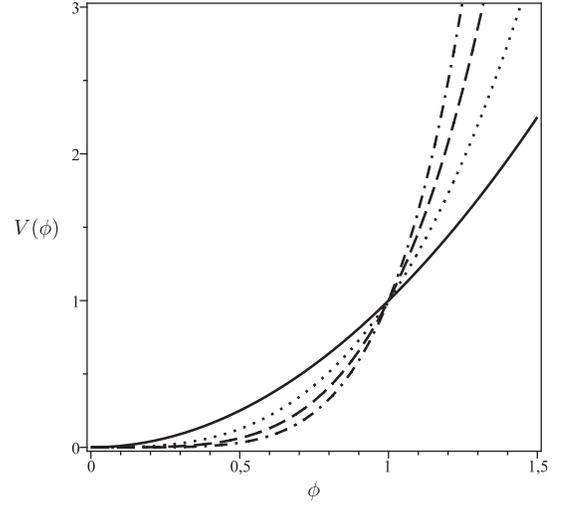}
\caption{Plots of the chaotic inflation potential (solid line) and of the ethernal inflation potential to $p=3$ (dotted line), $p=4$ (dashed line), and $p=5$ (dotted-dashed line).}
\end{figure}

\begin{figure}[htb!]
\includegraphics[scale=.35]{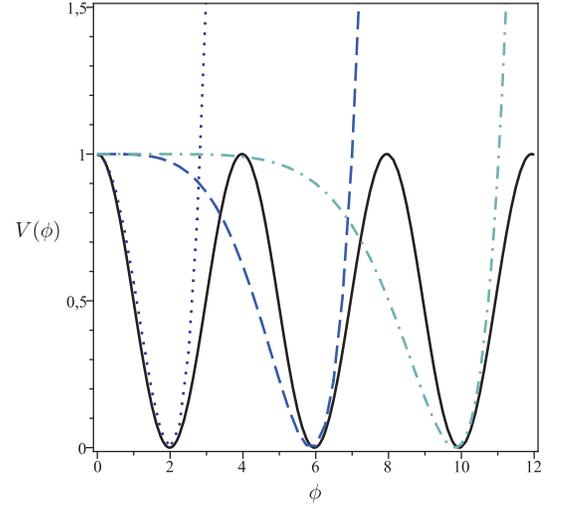}
\caption{Plots of the natural inflation potential (solid line) and of the hilltop inflation potential to $p=4$ (dotted line), $p=8$ (dashed line), and $p=12$ (dotted-dashed line). The parameter $\lambda$ was properly chosen so that the zeros of the natural inflation potential were coincident with the zeros of the hilltop inflation potential.}
\end{figure}

The e-fold number can be computed as
\be
N=\left[\chi^2\left(\frac1{p}+4\frac{V_0}{\lambda}\,\frac{\chi^{-p/2}}{p-4}\right)\right]^{\chi_{\text{end}}}_{\chi_{\text{ini}}}
\ee
and for the deformed model we have 
\be
\tilde{N}=\frac1{r^2}\ln\left(\frac{\sin(r\phi_{\text{ini}})}{\sin(r\phi_{\text{end}})}\right)
\ee
For $p=4$ we come to
\be
f(\phi)=\chi=\left(\frac{\sin(r\phi)}{1+\cos(r\phi)}\right)^{\pm\lambda\big/\left(2\sqrt{\tilde{V}_0}r^2\right)}
\ee
and the integration constant is such that $f(\phi=\pi/(2r))=1$. The slow-roll parameters $\epsilon$ and $\eta$ are the same as before above, with $p=4$ in their respective expressions.

The e-fold number is now
\be
N=\left[\frac{\chi^2}{4}-2\,\frac{V_0}{\lambda}\,\ln{\chi}\right]^{\chi_{\text{end}}}_{\chi_{\text{ini}}}
\ee

In both cases we can see that the parameters in the slow-roll inflation regime of the new model are constructed, through the deformation procedure, considering known results of the original model. We take special deformation functions only to illustrate the procedure, exploiting well-established results in the literature. However, if we take the choice of arbitrary deformation functions, we can generate new potentials, expanding the analytical range of solutions in the slow-roll inflation regime.   

\section{Summary} \label{sec6}

In this paper we have shown as the deformation procedure works on cosmological background, most especially in the slow-roll regime of the inflationary phase, since for this situation the first-order differential equations favor the application of the method. This does not impose a limit to the procedure, because we can reduce the order of the higher order differential equations, which is a direct way to improve the method for further analysis in the cosmological background, provided that a scalar field drives the evolution of the universe, that will be the focus of our interest on the current study in future papers.\\

The authors would like to thank CNPq Brasil and CAPES for partial support.


\end{document}